RESEARCH ARTICLE

# Image-Based Modeling of Blood Flow and Oxygen Transfer in Feto-Placental Capillaries

Philip Pearce[1]*, Paul Brownbill[2,3], Jiří Janáček[4], Marie Jirkovská[5], Lucie Kubínová[4], Igor L. Chernyavsky[1], Oliver E. Jensen[1]

**1** School of Mathematics, University of Manchester, Manchester, M13 9PL, United Kingdom, **2** Maternal and Fetal Health Research Centre, Institute of Human Development, University of Manchester, Manchester, United Kingdom, **3** Maternal and Fetal Health Research Centre, St. Mary's Hospital, Central Manchester University Hospitals NHS Foundation Trust, Manchester Academic Health Science Centre, Manchester, M13 9WL, United Kingdom, **4** Department of Biomathematics, Institute of Physiology, v.v.i., Academy of Sciences of the Czech Republic, Prague, Czech Republic, **5** Institute of Histology and Embryology, First Faculty of Medicine, Charles University, Albertov 4, CZ-12801 Prague 2, Czech Republic

* ppearce@mit.edu

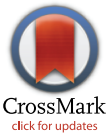

## Abstract

During pregnancy, oxygen diffuses from maternal to fetal blood through villous trees in the placenta. In this paper, we simulate blood flow and oxygen transfer in feto-placental capillaries by converting three-dimensional representations of villous and capillary surfaces, reconstructed from confocal laser scanning microscopy, to finite-element meshes, and calculating values of vascular flow resistance and total oxygen transfer. The relationship between the total oxygen transfer rate and the pressure drop through the capillary is shown to be captured across a wide range of pressure drops by physical scaling laws and an upper bound on the oxygen transfer rate. A regression equation is introduced that can be used to estimate the oxygen transfer in a capillary using the vascular resistance. Two techniques for quantifying the effects of statistical variability, experimental uncertainty and pathological placental structure on the calculated properties are then introduced. First, scaling arguments are used to quantify the sensitivity of the model to uncertainties in the geometry and the parameters. Second, the effects of localized dilations in fetal capillaries are investigated using an idealized axisymmetric model, to quantify the possible effect of pathological placental structure on oxygen transfer. The model predicts how, for a fixed pressure drop through a capillary, oxygen transfer is maximized by an optimal width of the dilation. The results could explain the prevalence of fetal hypoxia in cases of delayed villous maturation, a pathology characterized by a lack of the vasculo-syncytial membranes often seen in conjunction with localized capillary dilations.

## Introduction

Oxygen is important for both fetal and placental development throughout pregnancy [1]. Much of the research relating the structure of the placenta to fetal oxygen uptake has taken a stereological approach, using two-dimensional histological slides to calculate structural

**Data Availability Statement:** All relevant data are within the paper and its Supporting Information files.

**Funding:** P.P. was supported by an Engineering and Physical Sciences Research Council (EPSRC) Doctoral Prize Fellowship (EP/L504877/1). O.E.J. was supported by EPSRC, grant EP/K037145/1. I. L.C. was supported by Medical Research Council (MRC), grant MR/N011538/1. J.J. and L.K. were supported by the Czech Academy of Sciences (RVO: 67985823). M.J. was supported by the project PRVOUK P25/LF1/2 of Charles University in Prague. This work was supported by the EPSRC Mathematical Sciences Platform, Grant EP/I01912X/1. The funders had no role in study design, data collection and analysis, decision to publish, or preparation of the manuscript.

**Competing Interests:** The authors have declared that no competing interests exist.

quantities such as average volumes, surface areas and lengths [2]. Recently, three-dimensional (3D) representations of the feto-placental vasculature, obtained by reconstructing confocal laser scanning microscope images, have become available [3, 4], enabling more detailed analysis of structures in specific regions of the placenta. In this paper, we use mathematical modeling in combination with 3D images to investigate blood flow and oxygen transfer in feto-placental capillaries.

The feto-placental vasculature is contained inside villous trees, which bathe in maternal blood, allowing the transfer of nutrients without mixing of the two blood supplies. As fetal blood flows from the umbilical artery to the umbilical vein, it passes through a series of vessels of varying spatial scale. At the smallest scale are the capillaries in the terminal villi, bulbous structures that are thought to be the main sites for the transfer of oxygen from mother to fetus in the final trimester of pregnancy [5]. As the primary functional unit of the mature placenta, terminal villi and the fetal capillaries therein merit careful investigation.

A specific characteristic of fetal capillaries in the terminal villi that we focus on in this study is the presence of localized dilations. Such dilations, often called sinusoids, have been observed in numerous studies [5]. The dilations are often seen in conjunction with a vasculo-syncytial membrane, where the capillary bulges against the villous surface. Due to organelle displacement, in these regions the maternal and fetal blood are separated only by a simple bilayer consisting of a capillary endothelium and a thin syncytiotrophoblast. Diverse physical explanations for the localized dilations have been hypothesized, including a reduction in feto-maternal diffusion distance, a reduction in the total resistance of the feto-placental vasculature and a deceleration in local blood flow to facilitate exchange [5].

There are a few sources linking vasculo-syncytial membranes, and the corresponding localized fetal capillary dilations, to oxygen transfer. First, there is thought to be an increased amount of fetal capillary dilation in pregnancies at high altitude, where conditions are hypoxic [6], evidenced by a reduced mean trophoblast thickness [7] and an increased percentage of villous volume occupied by fetal capillaries [8]. Second, a lack of terminal villi and vasculo–syncytial membranes, amongst other markers, characterizes a pathology known as delayed villous maturation, also known as defective villous maturation or distal villous immaturity [9–14]. There is an increased prevalence of delayed villous maturation in stillbirths where the autopsy findings were consistent with fetal hypoxia [10, 15]; it is also associated with poor neurologic outcomes due to hypoxia [9, 13]. There is thought to be an approximately 5% risk of delayed villous maturation recurring in subsequent pregnancies [15].

Past efforts to model the placenta theoretically can generally be split into those that model the maternal side with assumptions about the fetal side, and *vice versa*; extensive reviews of the historical literature are available in [16, 17]. In recent years, maternal flow has been modeled using Darcy's law for flow in a porous medium [16], as a uniform flow field [18, 19], or using the Navier–Stokes equations in combination with images of fetal villous trees [20]. Where applicable, in these studies oxygen transfer to the fetal vasculature has been simplified, either by including a reaction term in the advection–diffusion equation [16], or by assuming fetal vessels to be oxygen sinks [18, 19, 21].

Oxygen diffusion from maternal to fetal blood within villous branches has been simulated in realistic 2D geometries from histology slides [22] and 3D geometries reconstructed from confocal laser scanning microscopy [23], with fetal blood vessels assumed to be perfect oxygen sinks in both cases. This assumption corresponds to the situation where all of the provided oxygen is carried away by the fetal blood, and the oxygen transfer rate in this case is expected to provide an upper bound on oxygen transfer to the fetal capillary when blood flow is taken into account.

Attempts to model blood flow in the feto-placental vasculature have simplified the microstructure in the terminal villi [24]; a problem is how to capture the microstructure in the smallest feto-placental capillaries accurately. One solution is to simulate the shape and structure of the placenta and the fetal vessels it contains [24–26]. However, to improve on existing models it is vital to understand characteristics such as the vascular resistance of real capillaries in the feto-placental vasculature. This is particularly important when modeling the transfer of a substance such as oxygen, which can be dominated by advection (due to oxygen's affinity for hemoglobin in red blood cells) and is therefore significantly influenced by patterns of blood flow through capillaries. Spatial characteristics of oxygen transfer are also expected to be affected by the extent to which the oxygen content of the fetal blood has equilibrated with that of the maternal blood.

A multi-scale mathematical model of the entire placenta, spanning spatial scales from the umbilical cord to the terminal villi, would bring significant benefits. First, it would help identify the physical reasons for observed structures in healthy and diseased placentas. Second, it could aid clinicians in early detection and prevention of pathologies. Finally, it could enable patient-specific diagnostics using future imaging technologies, improving personalized medicine during pregnancy. This is an ambitious goal that will demand integration of contributions at many levels of spatial organization.

Three-dimensional images of fetal capillaries can be a useful tool for building theoretical models that reflect the microstructure of the placenta accurately. Because it is currently not feasible to solve the equations for blood flow and oxygen transfer in realistic geometries throughout an entire placenta, the equations can be solved in a selection of geometries to find statistical distributions of important physical properties. Such simulations are performed in the present study, using 3D images of fetal villous and capillary surfaces, albeit for a small number of examples. Several key properties that will be useful in future multi-scale models are calculated, including the vascular resistance and the total oxygen transfer rate.

It is important to be able to quantify the effects of statistical variability, experimental uncertainty and pathological placental structure on the generated statistical distribution of physical properties. This paper demonstrates two separate methods that can be used to quantify such effects. In the first method, scaling arguments are used to make predictions and assess the sensitivity of the model to changes in the geometry and the parameters. Three separate physical regimes are identified, which capture the oxygen transfer across a range of Péclet numbers: a diffusion-limited regime where oxygen transfer is determined by diffusion through the surrounding villous volume, and two flow-limited regimes, where oxygen transfer is determined by the pressure drop through the capillary and the vascular resistance. These physical regimes, and the associated scalings for the oxygen transfer, are summarized in Table 1. In the second method, an idealized axisymmetric model is used to quantify the influence of localized feto-placental capillary dilations, or a lack thereof, on oxygen transfer. As well as furthering our understanding of the effect of pathologies on blood flow and oxygen transfer, the results, which are validated by comparison with the 3D image-based modeling, provide physiological insight to help to prevent, identify and treat delayed villous maturation.

## Methods

### Image acquisition, mesh conversion and skeletonization

Three-dimensional (3D) images of fetal capillaries and villous surfaces were used from previous studies, in which the methods for specimen preparation, image acquisition and 3D image reconstruction are described in detail [3, 4]. All tissue was acquired with written consent from fully-informed women attending the General University Hospital in Prague for their antenatal

Table 1. Summary of oxygen transfer scaling results for $Pe_{eff} = Bu_0R_0/D \gg 1$.

| Oxygen transfer scaling | Equation | Expected validity |
| --- | --- | --- |
| Flow-limited scaling ($N_{flow,1}$) | Eq (12) | $Pe_{eff} \gg LR_0/(d+R_0)^2$, $N_{flow} \ll N_{max}$ |
| Flow-limited scaling ($N_{flow,2}$) | Eq (13) | $Pe_{eff} \ll LR_0/(d+R_0)^2$ |
| Diffusion-limited scaling ($N_{max}$) | Numerical (or approximation Eq (14)) | $N_{flow} \gg N_{max}$ |
| Regression equation | Eq (19) | All regimes in this study ($Pe_{eff} \gg 1$) |

The numerical result $N_{flow}$ can be calculated by assuming oxygen to be provided directly at the capillary surface. Here $Pe_{eff}$ is the effective Péclet number, $B$ is the oxygen advection enhancement parameter, $u_0$ is a typical flow speed (which is proportional to the pressure drop $\Delta P$), $R_0$ is the capillary radius, $D$ is the diffusion coefficient, $L$ is the capillary length and $d$ is the villous thickness. The scaling Eq (12) is expected to be valid when the villous volume is not too thick (i.e. when $d$ is of the same order of magnitude as $R_0$ or smaller).

doi:10.1371/journal.pone.0165369.t001

care. Tissues were collected under ethical approval by the Ethics Committee of the General University Hospital in Prague. Samples from healthy placentas were taken following uniform random sampling no later than 10 min after delivery. The samples were fixed in 4% formaldehyde containing 0.5% eosin for at least 24 h, and embedded in paraffin wax; thick sections (120 $\mu$m) were cut. Stacks of optical sections 1 $\mu$m apart were imaged using a Biorad MRC 600 laser scanning confocal microscope. Surfaces of villi and capillaries were created and rendered using the image analysis program *Ellipse* and converted into a 3D binary image. The image was resampled as a grey image with lower resolution and treated by 3D Gaussian filtration. The triangulated isosurface of the grey value was detected.

Triangulated isosurfaces were reconstructed using a Poisson surface reconstruction algorithm in the open-source mesh processing package *Meshlab* (version 1.3.3, http://meshlab.sourceforge.net) to form a smoother surface. The surfaces were imported into *Comsol Multiphysics* (version 5.2, http://comsol.com) and converted into 3D tetrahedral meshes using the inbuilt meshing algorithm. In the process, the surface mesh was further simplified to remove small defects. Geometries were partitioned with a plane to provide a flat surface for inflow and outflow boundary conditions to be applied (Fig 1A).

For skeletonization, triangulated isosurfaces were converted into tetrahedral meshes in the open-source mesh generating package *Gmsh* (version 2.12, [28]) and imported into the image analysis software *Avizo* (Lite version 9.0, https://www.fei.com/software/avizo3d). The capillaries were skeletonized using the *Avizo* Skeletonization Pack (Fig 1B). This uses a distance-ordered thinning algorithm to remove voxels successively until a skeleton with the thickness of a single voxel remains; the algorithm is described in detail in [29]. The total length of each capillary was defined as the total length of the skeletonization line, including all branches. At each point along the skeletonization line, the radius of the capillary was defined as the distance between the voxel on the line and the nearest capillary surface voxel. At each point on the capillary surface, the villous distance was defined as the distance between the capillary surface voxel and the nearest villous surface voxel. The unpartitioned geometry was used for the calculation of the mean and standard deviation of the villous distance. Further details of the methods used for image analysis are provided in S1 Appendix.

### Simulation of blood flow and oxygen transfer in 3D geometries

The parameters used in the simulations are summarized in Table 2. Blood flow in feto-placental capillaries has low Reynolds number, and was therefore modeled using the Stokes equations

$$\nabla \cdot \mathbf{u} = 0, \quad \nabla p = \mu \nabla^2 \mathbf{u}, \qquad (1)$$

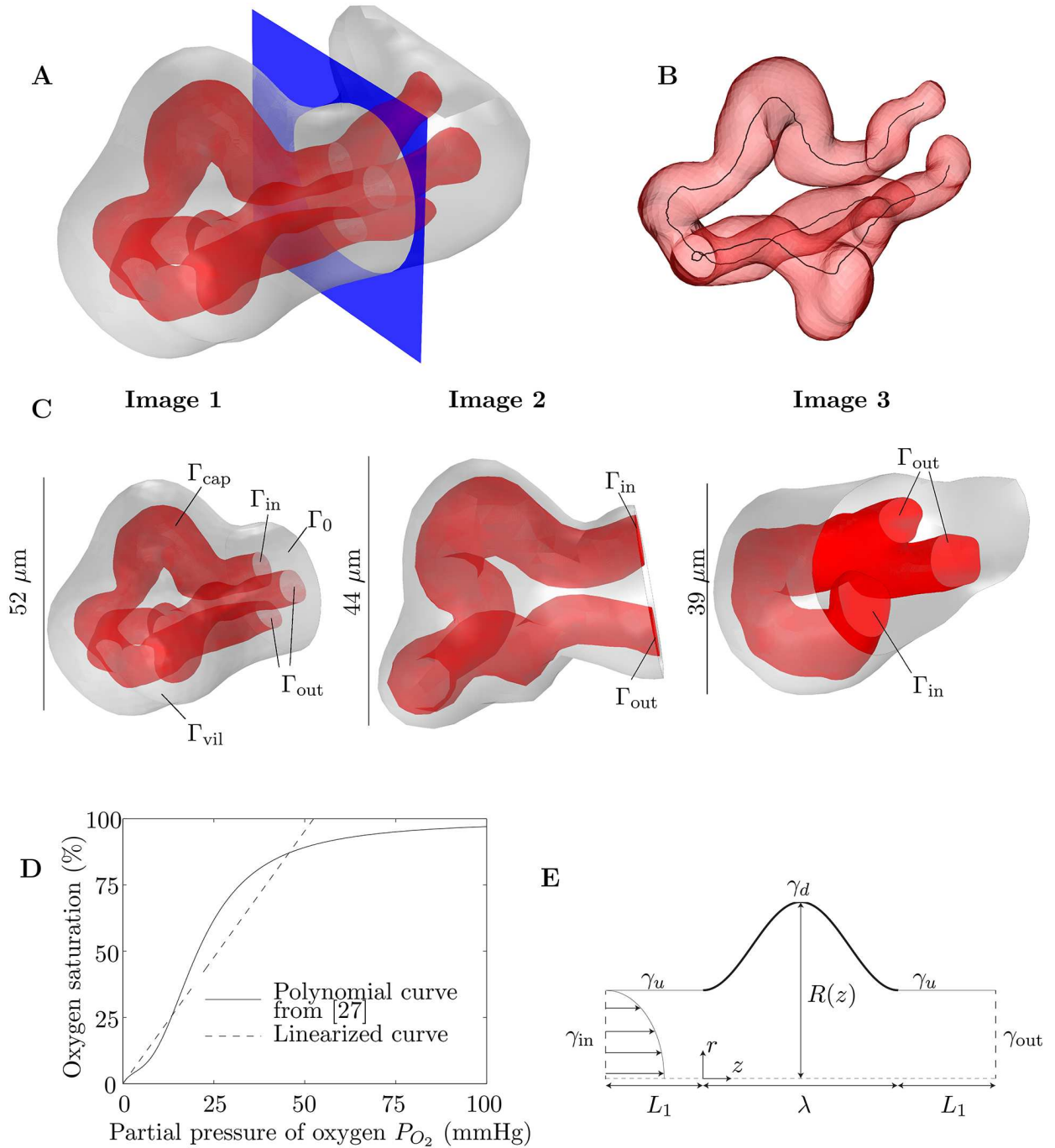

**Fig 1. Methods.** (A) An example of how 3D images are partitioned with a plane (shown in blue) to provide a surface for boundary conditions to be applied. The red surface represents the capillary surface (endothelium, $\Gamma_{cap}$) and the grey surface represents the villous surface $\Gamma_{vil}$. In this example the capillary bifurcates within the villous branch. (B) Example of the skeletonization of a capillary from a 3D image. The skeletonization line, representing the centreline of the lumen, is shown in black. (C) The 3D images of fetal capillaries and villous surfaces used in the simulations (see S1 Images). The images shown have been partitioned by a plane on which boundary conditions are applied, for inflow ($\Gamma_{in}$), outflow ($\Gamma_{out}$) and villous tissue ($\Gamma_0$). (D) Polynomial fetal oxygen–hemoglobin dissociation law from [27] (solid line) and the linear approximation between 0 and 60 mmHg, passing through the origin, found using a least-squares fit (dashed line). The gradient of the linear approximation is $K$ = 0.019 mmHg$^{-1}$. (E) Schematic diagram of the idealized axisymmetric model of a fetal capillary dilation, showing an axisymmetric tube with a localized dilation. Blood flows into the capillary through $\gamma_{in}$ and leaves through $\gamma_{out}$. Oxygen is provided along the dilated section of the capillary $\gamma_d$, denoted by the thicker black line. The undilated sections of the capillary are labeled $\gamma_u$.

doi:10.1371/journal.pone.0165369.g001

Table 2. Literature-based and calculated parameters used in the formulation of the problem and the estimation of oxygen advection enhancement by fetal hemoglobin.

| Parameter | Value | Reference |
|---|---|---|
| Diffusion coefficient of oxygen in plasma ($D$) | $1.7 \times 10^3$ $\mu m^2/s$ | [31] |
| Concentration of dissolved oxygen in the intervillous space ($c_{mat}$) | 0.07 mol/m$^3$ (2.24 g/m$^3$) | |
| Dynamic viscosity of plasma ($\mu$) | $1 \times 10^{-3}$ Pa s | [32] |
| Oxygen advection enhancement parameter ($B$) | 141 | S2 Appendix |

doi:10.1371/journal.pone.0165369.t002

where **u** is the flow velocity field, $p$ is the pressure and $\mu$ is the dynamic viscosity. For simplicity, flow was assumed to be steady and non-Newtonian effects of red blood cells (RBCs) on the flow were ignored; the suspension was assumed to have constant effective viscosity $\mu$. At small scales, the viscosity of blood is known to depend on the radius of the capillary it is flowing through, via the Fahraeus–Lindqvist effect [30]. The dynamics of blood flow in a capillary of varying radius, however, are not clear. Incorporating the effect of RBCs in 3D geometries like the ones studied in this paper would necessitate computations involving discrete RBCs, which is beyond the scope of the study.

Oxygen transfer in villous branches was modeled by the advection–diffusion equation, with oxygen transfer outside capillaries assumed to be by diffusion alone. The effect of oxygen binding to hemoglobin in blood was taken into account using the method described in [18], which is summarized briefly here.

Blood is assumed to be a continuum, with hemoglobin and plasma occupying the whole volume, a reasonable assumption at normal hematocrits [33]. The total oxygen concentration is split into the concentration of oxygen dissolved in plasma $c$ (per unit volume of blood) and the concentration of oxygen bound to hemoglobin $c_h$ (per unit volume of blood). Oxygen is assumed to diffuse only when dissolved in the plasma due to the low diffusion coefficient of RBCs [34], to give

$$\mathbf{u} \cdot \nabla (c + c_h) = D \nabla^2 c, \qquad (2)$$

where $D$ is the diffusion coefficient of oxygen in the plasma (assumed to be constant throughout the villous branch). Oxygen is assumed to bind to hemoglobin instantaneously, which can be justified by comparing the typical time for oxygen–hemoglobin dissociation ($\tau_d \approx 20$ ms for the slowest process [35]) to the typical transit time of blood through a capillary ($\tau_t \approx 0.3$ s for a capillary of length 100$\mu$m, with centreline blood flow velocity of 300 $\mu$m/s [36]). The concentration of oxygen bound to hemoglobin is related to the partial pressure of oxygen in the plasma $P_{O_2}$ by the oxygen–hemoglobin dissociation law

$$c_h = c_{\max} S(P_{O_2}), \quad P_{O_2} = \frac{k_{hn}}{\rho_{bl}} c. \qquad (3)$$

Here $c_{\max}$ is the oxygen content of fetal blood at full saturation and the partial pressure of oxygen in the plasma has been written using Henry's law, where $k_{hn}$ is the Henry's law coefficient and $\rho_{bl}$ is the density of blood. Inserting Eq (3) into Eq (2) gives

$$\left(1 + \frac{c_{\max} k_{hn}}{\rho_{bl}} S'(P_{O_2})\right) \mathbf{u} \cdot \nabla c = D \nabla^2 c. \qquad (4)$$

Finally, the fetal oxygen–hemoglobin dissociation law given in [27] is linearized in the region between 0 mmHg and 60 mmHg (where 60 mmHg is the highest partial pressure of oxygen the fetal blood is expected to be exposed to [37]) using a least-squares fit, with the constraint that the curve passes through the origin (Fig 1D), so that

$$S'(P_{O_2}) = KP_{O_2}, \tag{5}$$

for $K = 0.019$ mmHg$^{-1}$. Inserting Eq (5) into Eq (4) gives

$$B\mathbf{u} \cdot \nabla c = D\nabla^2 c, \tag{6}$$

where $B = 1 + c_{max}Kk_{hn}/\rho_{bl}$ is a parameter quantifying the enhancement to advection due to oxygen's affinity for hemoglobin.

Using typical values of the parameters, it was estimated that $B \approx 141$ (see S2 Appendix). The uptake of oxygen by the placenta itself, via the outer layer of the villous surface (the syncytiotrophoblast), was neglected. The placenta is thought to consume approximately 40% of the oxygen leaving the maternal circulation [38]. However, it is not known what proportion of this consumption takes place in the terminal villi.

Surfaces where boundary conditions were applied are labeled in Fig 1C. The pressure difference between the inflow boundary $\Gamma_{in}$ and the outflow boundary $\Gamma_{out}$ was fixed (for flow in a rigid domain, pressure differences, rather than absolute pressures, are relevant). As is common in flow simulations [39], the normal stress was specified at these boundaries, with the tangential velocity vector $\mathbf{u}_t$ set to zero. Thus the inlet pressure was fixed at $p = \Delta P$, say, by specifying the normal stress $\mathbf{n} \cdot \boldsymbol{\sigma} \cdot \mathbf{n} = -\Delta P$, where $\boldsymbol{\sigma} = -p\mathbf{I} + \mu(\nabla\mathbf{u} + (\nabla\mathbf{u}^T))$ is the stress tensor, $\mathbf{I}$ is the identity tensor and $\mathbf{n}$ is the outer unit normal to the boundary. A no-slip condition was applied on the capillary wall $\Gamma_{cap}$, neglecting the effect of the endothelial glycocalyx. In summary, the boundary conditions applied to the velocity field were

$$\mathbf{n} \cdot \sigma \cdot \mathbf{n} = -\Delta P, \quad \mathbf{u}_t = \mathbf{0} \quad \text{on } \Gamma_{in}, \tag{7a}$$

$$\mathbf{n} \cdot \sigma \cdot \mathbf{n} = 0, \quad \mathbf{u}_t = \mathbf{0} \quad \text{on } \Gamma_{out}, \tag{7b}$$

$$\mathbf{u} = \mathbf{0} \quad \text{on } \Gamma_{cap}. \tag{7c}$$

At the inflow boundary $\Gamma_{in}$, the blood was considered to be deoxygenated. At the villous surface $\Gamma_{vil}$, the concentration of oxygen was assumed to be equal to the concentration $c_{mat}$ of oxygen in the maternal blood in the intervillous space. There is evidence that this is reasonable from previous studies showing the maternal flow to have spatial gradients over the lengthscale of an entire placentone [16, 20]. To ensure the simulations were performed in the range of partial pressures of oxygen where the linearization of the oxygen–hemoglobin dissociation law is valid (Fig 1D), a value of $c_{mat} = 0.07$ mol/m$^3$ was taken, corresponding to a partial pressure of oxygen in the maternal blood of 52.5 mmHg. For larger partial pressures, the effect of saturation of hemoglobin may need to be taken into account. Zero diffusive flux conditions were applied at the outflow boundary $\Gamma_{out}$ and the surface $\Gamma_0$. In summary, the boundary conditions applied to the concentration were

$$c = 0 \quad \text{on } \Gamma_{in}, \quad c = c_{mat} \quad \text{on } \Gamma_{vil}, \quad \frac{\partial c}{\partial n} = 0 \quad \text{on } \Gamma_{out} \text{ and } \Gamma_0, \tag{7d}$$

where $n$ is the coordinate normal to the boundary.

Branched capillaries (Image 1 and Image 3, Fig 1C) were assumed to have one inflow and two outflow boundaries, with blood assumed to flow in the direction where the branching angle was smallest. In the unbranched capillary (Image 2, Fig 1C), $\Gamma_{in}$ and $\Gamma_{out}$ were designated arbitrarily; it was found in computations that reversing $\Gamma_{in}$ and $\Gamma_{out}$ led to less than 0.01%

Table 3. Geometric and calculated properties of the three 3D images used in the paper.

| Geometric property | Image 1 | Image 2 | Image 3 |
|---|---|---|---|
| Villous Volume ($\mu m^3$) | $6.88 \times 10^4$ | $5.62 \times 10^4$ | $2.54 \times 10^4$ |
| Villous Surface Area ($\mu m^2$) | $8.01 \times 10^3$ | $7.55 \times 10^3$ | $3.96 \times 10^3$ |
| Capillary volume as proportion of villous volume | 18.5% | 34.0% | 25.5% |
| Capillary surface area as proportion of villous surface area | 68.0% | 86.2% | 67.4% |
| Total capillary length $L$ ($\mu m$) | 185 | 164 | 92 |
| Longest path from inflow to outflow boundary ($\mu m$) | 147 | 164 | 80 |
| Mean villous distance from capillary surface $d$ (± s.d.) ($\mu m$) | 51 (±19) | 15 (±10) | 22 (±11) |
| **Calculated property** | | | |
| Maximum velocity magnitude $u_{max}$ at $\Delta P = 0.5$ Pa ($\mu m/s$) | 45 | 38 | 53 |
| Vascular resistance $R = \Delta P/Q$ (Pa s/$\mu m^3$) | $5.8 \times 10^{-4}$ | $4.0 \times 10^{-4}$ | $3.0 \times 10^{-4}$ |
| Ratio $L^2/R$ for scaling predictions ($\mu m^5$/Pa s) | $5.9 \times 10^7$ | $6.7 \times 10^7$ | $2.8 \times 10^7$ |
| Vascular resistance per unit capillary volume $R_{cap}$ (Pa s/$\mu m^6$) | $4.58 \times 10^{-8}$ | $2.1 \times 10^{-8}$ | $4.7 \times 10^{-8}$ |
| Vascular resistance per unit villous volume $R_{vil}$ (Pa s/$\mu m^6$) | $8.4 \times 10^{-8}$ | $7.1 \times 10^{-9}$ | $1.2 \times 10^{-8}$ |
| Upper bound on oxygen transfer rate $N_{max}$ ($\mu g/s$) | $3.2 \times 10^{-6}$ | $6.6 \times 10^{-6}$ | $2.7 \times 10^{-6}$ |
| Theoretical prediction of $N_{max}$ from Eq (14) ($\mu g/s$) | $1.8 \times 10^{-6}$ | $3.3 \times 10^{-6}$ | $1.2 \times 10^{-6}$ |
| Oxygen transfer ratio $N_{ratio} = N/N_{flow}$ at $\Delta P = 0.5$ Pa | 83% | 90% | 84% |
| Oxygen transfer ratio $N_{ratio} = N/N_{flow}$ at $\Delta P = 20$ Pa | 28% | 39% | 26% |

The geometric properties correspond to the images after partitioning by a plane for boundary conditions to be applied. The maximum velocity magnitude at $\Delta P = 0.5$ Pa can be used to calculate the value at any other pressure drop by multiplying by the proportional change in $\Delta P$.

doi:10.1371/journal.pone.0165369.t003

difference in the flow rate (to be expected from the reversibility of Stokes flow) and less than 0.1% difference in the total oxygen transfer rate.

The equations were solved numerically in *Comsol Multiphysics* using the direct solver MUMPS [40, 41]. Solutions were computed on dual E5-2643v3 CPU nodes with hyper-threading disabled and 768GB of RAM. Results were checked to be independent of the mesh, which consisted typically of $\sim 10^6$–$10^7$ tetrahedral elements. Simulations were performed on three different 3D geometries (Fig 1C). Geometric characteristics of the corresponding capillary and villous surfaces are summarized in Table 3. The total oxygen transfer rate $N$ from the maternal blood was calculated by integrating the advective flux leaving the domain through $\Gamma_{out}$, which was checked to be equal to the diffusive flux entering the domain across $\Gamma_{vil}$ minus the diffusive flux leaving through $\Gamma_{in}$ (an artifact due to the imposed boundary condition, which does not affect the results significantly for flows in which advection dominates diffusion). The integral of the diffusive flux through each boundary was computed accurately by a summation of the corresponding components of the boundary residual vector over every node on the surface [42].

Due to the linearity of the Stokes equations, the flow field needs only one computation for each capillary to calculate the vascular resistance $R = \Delta P/Q$, where $\Delta P$ is the pressure drop between inflow and outflow and $Q$ is the flow rate. Similarly, a single solution of the advection–diffusion Eq (6) can be used to calculate the non-dimensional concentration field, $\hat{c} = (c - c_{in})/(c_{mat} - c_{in})$, where $c_{in}$ is the inflow oxygen concentration. Rearranging this equation gives the dimensional concentration field for any values of $c_{in}$ and $c_{mat}$. Further details of the methods use in the 3D simulations are given in S1 Appendix.

### Idealized axisymmetric model and optimization of capillary dilation

The parameters used in the axisymmetric model of an individual capillary dilation are summarized in Table 2. Blood flow and oxygen transfer were modeled by Eqs (1) and (6) in a simplified geometry consisting of an axisymmetric capillary of length $L$, with a dilated section of length $\lambda$ in the centre of the capillary wall (Fig 1E), where the undilated radius was $R_0$. To model a vasculo-syncytial membrane, oxygen was assumed to be provided only along the dilated section of the capillary, due to its proximity to the maternal blood; deoxygenated blood was assumed to flow into the capillary, driven by a pressure difference across the capillary $\Delta P$, which was fixed to give a centreline velocity of 300 $\mu$m/s in an undilated capillary [36]. A no-slip condition was applied on the capillary wall. The diffusive flux of oxygen was set to zero on the undilated walls and the outflow boundary. To summarize, the applied boundary conditions were

$$\mathbf{n} \cdot \sigma \cdot \mathbf{n} = -\Delta P, \quad \mathbf{u}_t = \mathbf{0}, \quad c = 0, \quad \text{on } \gamma_{in}, \quad (8a)$$

$$\mathbf{u} = \mathbf{0}, \quad c_r = 0, \quad \text{on } \gamma_u, \quad (8b)$$

$$\mathbf{u} = \mathbf{0}, \quad c = c_{\text{mat}}, \quad \text{on } \gamma_d, \quad (8c)$$

$$\mathbf{n} \cdot \sigma \cdot \mathbf{n} = 0, \quad \mathbf{u}_t = \mathbf{0}, \quad c_z = 0, \quad \text{on } \gamma_{out}, \quad (8d)$$

where the boundaries $\gamma_i$ are illustrated in Fig 1E. Arteriolar and venular resistances and oxygen transfer upstream and downstream of the capillary were ignored for simplicity (although a modification of condition Eq (8b) on $\gamma_u$ is discussed in S3 Appendix).

To reduce the degrees of freedom (DOF) in the optimization, the shape of the dilation $R(z)$ was expanded as a Fourier series truncated to $n$ terms,

$$\frac{R(z)}{R_0} = \frac{a_0}{2} + \sum_{k=1}^{k=n} a_k \cos(k\pi z/\lambda) + \sum_{k=1}^{k=n} b_k \sin(k\pi z/\lambda), \quad 0 < z < \lambda, \quad (9)$$

where $z$ is the axial coordinate. To ensure a continuous and smooth capillary wall, the conditions

$$R(0) = R(\lambda) = R_0, \quad R'(0) = R'(\lambda) = 0, \quad (10)$$

were imposed. The conditions Eq (10) provided constraints on the parameters $a_k$ and $b_k$. Further, to avoid errors in the simulations, the extrema of $R(z)$ were constrained between prescribed values; it was checked that the chosen values did not affect the optimal solutions. In cases where the shape was assumed to be symmetric about the axial midpoint of the dilation, $a_k$ and $b_k$ were set to zero for odd values of $k < n$.

Shape optimization was performed using the *fmincon* function in *Matlab*, which uses a Sequential Quadratic Programming algorithm [43]. No further regularization was imposed beyond the implicit regularization in the choice of $n$ (and in some cases the assumption of symmetry), which restricted the number of DOF in the shape of the dilation. At each step in the optimization process, the finite-element problem was solved numerically in *Comsol Multiphysics*, which output the total oxygen transfer rate $N$. The number of DOF in the optimization was increased successively.

The function *fmincon* found a local maximum in $N$ in the parameter-space of $a_0$, $a_k$ and $b_k$, $k < n$. An important aspect of the optimization was how to check that the local maximum in $N$ was in fact a global maximum. In cases with one DOF, the solution was checked to be optimal by comparing against a curve of N versus the maximum radius. In cases with two DOF, results were checked against contour plots of N in parameter space. In cases with successively increasing DOF, at least ten optimizations were performed, with initial conditions taken as a random sample in parameter space, constrained to ensure the dilation did not occlude the channel. If more than one local maximum was found, the shape with the largest oxygen transfer was taken to be the optimal solution. The non-invasive optimization method described here, using *Comsol Multiphysics* as a "black box" solver, was chosen for speed and simplicity.

When analyzing a dilation from a skeletonized capillary branch of length L, quantifying the geometry of a dilation required measuring the maximum radius $R_{max}$ and the undilated radius $R_0$, and defining the start and end points of the dilation to calculate its length $\lambda$. Two different definitions were used: the first defined the beginning and end of the dilation as local minima in the radius, with the undilated radius being the minimum radius of the capillary branch; the second used the average radius of the capillary branch as the undilated radius and defined the edges of the dilation where the radius passed its average radius. These two separate definitions provided a range of geometrical values to be used as input to the axisymmetric model.

## Results

### Blood flow and oxygen transfer in real geometries: scaling and simulations

The calculated values of the vascular resistance R, the vascular resistance per unit capillary volume $R_{cap}$ and the vascular resistance per unit villous volume $R_{vil}$ for each capillary in Fig 1C are given in Table 3. The capillaries with the least vascular resistance are those with branches, as expected; the total vascular resistance of two capillaries in parallel, each with the same resistance, is half the resistance of each capillary. As well as branching, the most important geometrical determinants of the vascular resistance are the average radius $R_0$ and the total length L of the capillaries; recall from Poiseuille's law that in a straight capillary with fully developed flow, R is given by

$$R = \frac{8\mu L}{\pi R_0^4}. \quad (11)$$

The vascular resistance of the capillaries can be used to predict which capillary is expected to receive the most oxygen from the maternal side. To this end, consider a straight capillary of radius $R_0$ and length L, with fixed concentration $c_{mat}$ at the wall and fully developed Poiseuille flow entering deoxygenated. The scalings determining the oxygen transfer rate depend on the magnitude of the effective Péclet number $Pe_{eff} = BR_0 u_0/D$, which gives the ratio of transport by advection to transport by diffusion, including the effect of oxygen transport by hemoglobin in red blood cells. Here D is a diffusion coefficient, $u_0$ is a characteristic flow speed, which is proportional to the pressure drop $\Delta P$ across the capillary, and B is a numerical factor that quantifies the enhancement to advection as a result of oxygen binding to hemoglobin (see Eq (6)). When $Pe_{eff} \gg 1$, advection dominates and, close to the capillary entrance, oxygen transfer is confined to a thin diffusive boundary layer near the capillary wall, where the axial component of the fluid velocity increases linearly with the distance from the wall and the curvature of the wall can be neglected [44]. Axial diffusion is expected to be of relative magnitude $\sim Pe_{eff}^{-2}$, and can therefore be neglected everywhere except in the region very close to the entrance [45], which we ignore for simplicity. The resulting boundary-layer equation can be solved with a similarity variable to find that the local oxygen flux per unit area along the capillary wall is proportional to $c_{mat}D(Pe_{eff}/(R_0^2 \, z))^{1/3}$, where z is the axial coordinate [44].

Integrating across the capillary surface and using Eq (11) gives the total oxygen transfer rate (labeled $N_{\text{flow}}$ as it corresponds to the flow-limited regime) to be

$$N_{\text{flow},1} \sim \alpha c_{\text{mat}} (D^2 B \Delta P L^2 / R)^{1/3}, \quad (12)$$

where $\alpha = (12\pi^2)^{1/3}/\Gamma(4/3) \approx 5.5$ [44]; here $\Gamma$ is the gamma function. The approximation Eq (12) is expected to hold for sufficiently short capillaries with $Pe_{\text{eff}} \gg 1$, as long as $Pe_{\text{eff}}$ is not large enough for oxygen transfer to be limited by diffusion through the villous volume. However, for the images in this study, if the ratio of capillary length $L$ to radius $R_0$ plus villous thickness $d$ is sufficiently large ($Pe_{\text{eff}} \ll LR_0/(d+R_0)^2$), oxygen is expected to diffuse to the centreline of the downstream region of the capillary, suppressing concentration gradients across the capillary cross-section [46] (if the simplifying assumption is made that capillary branches do not affect each other, the longest path between the inflow boundary and outflow boundary through a branched capillary could be used as $L$ in this inequality). In this case the concentration is said to be fully developed, or equilibrated, in the downstream region of the capillary. The amount of oxygen carried away by the flow is then proportional to the amount of blood flowing through the capillary,

$$N_{\text{flow},2} \sim c_{\text{mat}} B \Delta P / R. \quad (13)$$

Using the scalings Eqs (12) and (13), ignoring the effects of capillary branching and the space between capillary and villous surfaces for simplicity, it would be expected that for larger Péclet numbers, the greatest oxygen transfer would occur in Image 2 and the least in Image 3 (comparing $L^2/R$; see Table 3). For Péclet numbers small enough for the concentration to become equilibrated downstream, the greatest oxygen transfer would be expected to occur in Image 3 and the least in Image 1 (comparing $1/R$).

The predictions made above are confirmed by the numerical results (Fig 2A). At larger values of $\Delta P$, corresponding to $Pe_{\text{eff}} \sim 10^3$, oxygen transfer is found to be highest in Image 2 and lowest in Image 3, as expected. At smaller values of $\Delta P$, corresponding to $Pe_{\text{eff}} \sim 10^2$, oxygen transfer is found to be highest in Image 3 and lowest in Image 1, in line with the predicted behaviour. In this case, the oxygen distribution is insensitive to the flow of blood leaving the capillary (Fig 2C), and most of the oxygen transfer takes place at the capillary entrance and across the nearby villous surface (Fig 2B). For a higher pressure drop, oxygen transfer takes place across the entire capillary surface, although the largest amount of oxygen still enters around the capillary entrance and the nearby villous surface (Fig 2B and 2C). The majority of oxygen enters the capillary through the sections of the capillary wall closest to the villous surface (Fig 2B), illustrating the "diffusional screening" effect identified in [22, 47]. In fact, around the entrance to the villous branch, some oxygen leaves through the walls of the outflow capillaries furthest from the villous surface.

Another way to estimate which capillary is expected to receive the most oxygen for higher values of $\Delta P$ is by solving the diffusion equation in the villous volume surrounding the capillary, with the capillary surface assumed to be deoxygenated by setting $c = 0$ on $\Gamma_{\text{cap}}$, which is less computationally intensive than solving for flow in the capillary and avoids the need to resolve the boundary layers in the concentration that arise at larger values of $\Delta P$. The calculation, which is similar to those performed in [23], corresponds to the situation where all of the oxygen is carried away by the flow and oxygen transfer is limited by diffusion through the villous volume, providing an upper bound, $N_{\text{max}}$, on the total oxygen transfer rate $N$. The calculated numerical values of $N_{\text{max}}$ for each capillary are given in Table 3. The values of the total oxygen transfer per unit area $A$, given by $N_{\text{max}}/A$, are found to be of the same order of magnitude as the results of [23]. The value of $N_{\text{max}}$ can be related to the geometrical properties of

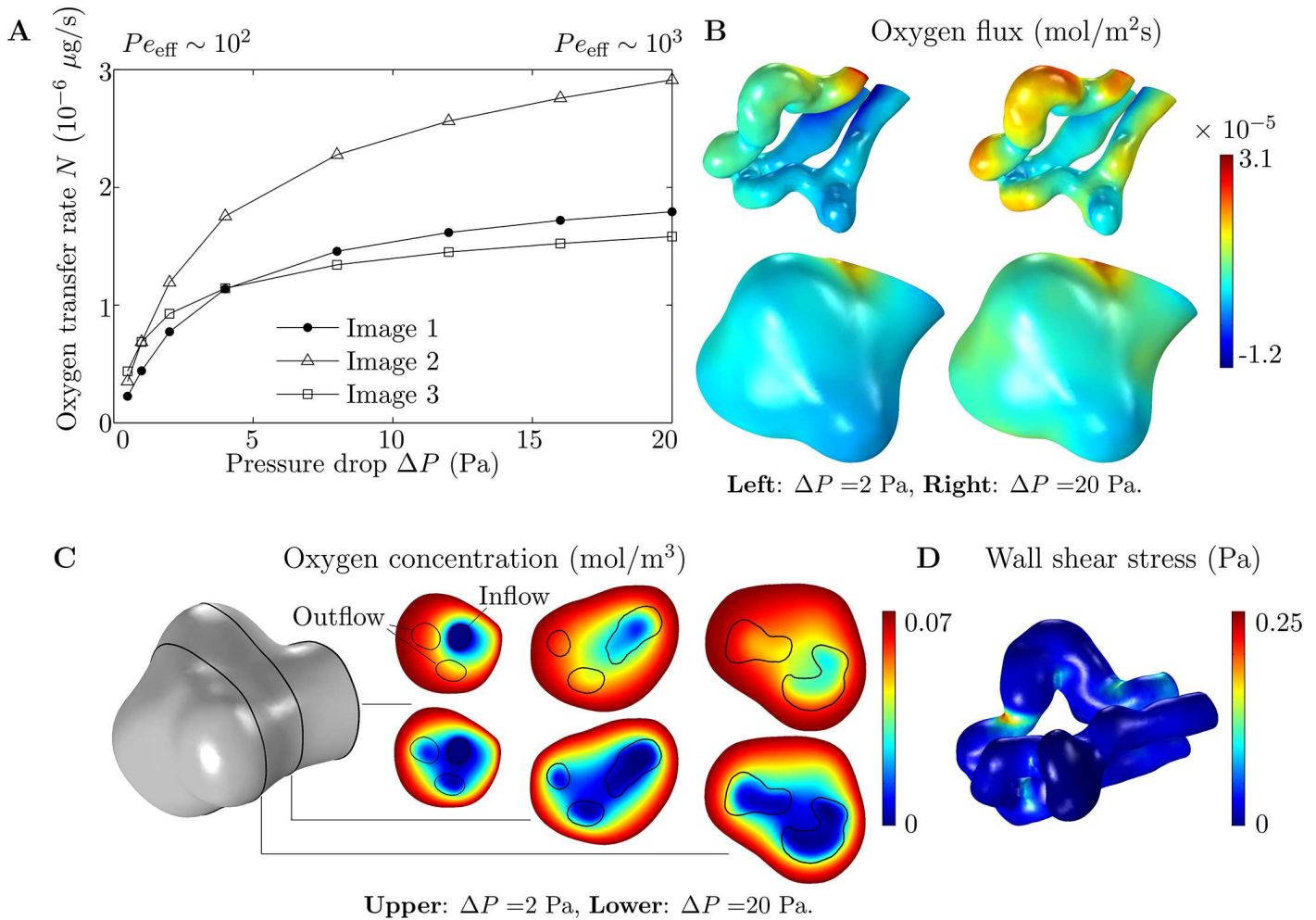

**Fig 2. Results of numerical simulations on 3D geometries.** (A) Oxygen flux entering each capillary versus pressure drop ΔP. The range of values of ΔP is broadly physiological, leading to flow velocities of around 300 μm/s (Table 3), which is the approximate flow velocity in normal capillaries [36]. $Pe_{eff}$ is in the range $10^2$–$10^3$ as indicated. (B) Distribution of oxygen flux entering through the capillary (top) and villous (bottom) surfaces, from simulations on Image 1. Results are shown for ΔP = 2 Pa (left) and ΔP = 20 Pa (right). (C) Plots of the oxygen concentration field on three planes, 15 μm apart, from simulations on Image 1. The planes are shown on the villous surface on the left. Results are given for ΔP = 2 Pa (top) and ΔP = 20 Pa (bottom). Capillary walls are denoted by black lines. The inflow and outflow capillaries are labeled on the plane at the villous entrance (left). (D) Wall shear stress at the capillary surface, from a simulation on Image 1 (ΔP = 2 Pa).

doi:10.1371/journal.pone.0165369.g002

each villous branch by considering a cylindrical capillary with a deoxygenated surface, with radius $R_0$ and surface area $A$, inside a cylindrical villous branch with an oxygenated surface, with radius $R_0 + d$. Solving the diffusion equation in the annulus between the concentric cylinders [44], the predicted maximum oxygen transfer rate is given by

$$N_{max} \approx \frac{D A \, c_{mat}}{R_0 \ln\left(1 + \frac{d}{R_0}\right)}. \tag{14}$$

Taking the radius of each capillary to be $R_0 = \sqrt{V/\pi L}$ (as for a cylindrical capillary of volume $V$ and length $L$), and using values of $V$, $L$ and $d$ from Table 3, Eq (14) gives the theoretical predictions of $N_{max}$ given in Table 3, which are of the same order of magnitude as the numerical results, and predict that Image 2 has the highest maximum oxygen transfer rate and Image 3 has the smallest, in agreement with the numerical results.

An oxygen transfer ratio $N_{\text{ratio}} = N/N_{\text{flow}}$ for each villous branch can be calculated by comparing oxygen transfer $N$ in the villous branch to oxygen transfer $N_{\text{flow}}$ in the situation where oxygen is provided directly at the surface of the capillary by imposing $c = c_{\text{mat}}$ on $\Gamma_{\text{cap}}$ (Fig 1C). There are expected to be two main factors influencing $N_{\text{ratio}}$. First, a capillary that is closer to the villous surface, on average, is expected to have a higher $N_{\text{ratio}}$ than a capillary that is further away, because the villous volume will have less effect on oxygen transfer in this case. Second, a capillary in which the concentration becomes more equilibrated is expected to have a higher $N_{\text{ratio}}$, because in this case the oxygen concentration at the capillary surface in the downstream region of the capillary is closer to that of the villous surface. Therefore $N_{\text{ratio}}$ should be higher at lower values of $\Delta P$. Image 2 is expected to have the highest $N_{\text{ratio}}$ because it has the highest value for the longest path length through its capillary, which is also the closest to the villous surface on average (Table 3). Images 1 and 3 are expected to have similar values of $N_{\text{ratio}}$ because the capillary in Image 1 is longer than the capillary in Image 3, but further away from the villous surface on average. As expected, Image 2 is found to have the highest $N_{\text{ratio}}$ at both $\Delta P = 0.5$ Pa and $\Delta P = 20$ Pa, while Images 1 and 3 are found to have roughly the same $N_{\text{ratio}}$ at both $\Delta P = 0.5$ Pa and $\Delta P = 20$ Pa (Table 3).

The average magnitude of the calculated capillary wall shear stress is found to vary between approximately 0.02 Pa and 0.25 Pa (or 0.2 dyn/cm$^2$ and 2.5 dyn/cm$^2$), which is close to what has recently been estimated to be exerted on the walls of the fetal villous trees by the maternal blood [20]. However, large spikes in wall shear stress are found around the inside of capillary bends, with dips around the outside of bends (Fig 2D). This could be considered counter-intuitive, as localized capillary dilations are thought to be important for the transfer of oxygen from the maternal blood, which would be enhanced by an increase in wall shear stress. A more thorough discussion of localized dilations can be found below.

It is important to understand the sensitivity of the model to changes in the geometry and the parameters. The scalings Eqs (12) and (13) suggest that, for a capillary of length $L$ and radius $R_0$, a uniform expansion (or contraction) of all lengthscales by magnitude $k$ would lead to an expected increase (or decrease) in the total oxygen transfer rate $N$ of $k^3$ for a smaller Péclet number (where the concentration becomes equilibrated in the capillary), or $k^{5/3}$ for a larger Péclet number (where the boundary-layer equation is valid, using the fact that the vascular resistance $R$ is proportional to $L/R_0^4$ from Eq (11)). A uniform expansion of 10% in all lengthscales ($k = 1.1$), which would be expected to cause an increase in the flow rate $Q$ of 33%, would therefore be expected to cause an increase in $N$ of around 33% for a smaller Péclet number or 17% for a larger Péclet number. Indeed, our computations show that applying a 10% uniform expansion to Image 2 leads to the expected 33% increase in $Q$, with a 33% increase in oxygen transfer at $\Delta P = 0.5$ Pa and a 15% increase at $\Delta P = 20$ Pa. A uniform expansion (or contraction) by magnitude $k$ in the capillary radius $R_0$ alone would be expected to have a slightly different effect, causing an increase (or decrease) in $N$ of $k^4$ for a smaller Péclet number or $k^{4/3}$ for a larger Péclet number. The sensitivity of the model to the parameters can also be assessed using the scalings Eqs (12) and (13) with the Formula (11) for the vascular resistance $R$.

### Regression equation for the total oxygen transfer rate

A general regression equation relating $N$ to $\Delta P$ across parameter space where $Pe_{\text{eff}} \gg 1$ should take into account the scaling Eq (13) at lower values of $\Delta P$, the scaling Eq (12) at higher values of $\Delta P$ and the upper bound on the oxygen transfer rate $N_{\text{max}}$, which are summarized in Table 1. A regression equation satisfying the scalings Eqs (12) and (13) in the case where oxygen is provided directly at the capillary surface is

$$N_{\text{flow}} \approx \frac{K_1 K_2 \Delta P}{K_2 + K_1 \Delta P^{2/3}}, \qquad (15)$$

which is similar to an equation regularly used to represent the results of experiments on heat transfer in tubes [48]. Here $K_1 = c_{\text{mat}} B/R$ and $K_2 = \alpha c_{\text{mat}} (D^2 BL^2/R)^{1/3}$. A correction to Eq (15) to account for the effect of the surrounding villous volume can be calculated by considering the diffusion of oxygen between a cylindrical villous surface of radius $R_0 + d$ surrounding a cylindrical capillary of radius $R_0$, with both cylinders of equal length. The oxygen concentration is fixed on the surface of the outer cylinder, so that

$$c = c_{\text{mat}} \text{ at } r = R_0 + d. \qquad (16)$$

The oxygen transfer rate per unit area is fixed on the inner cylinder, so that

$$D \frac{\partial c}{\partial r} = \frac{N}{A} \text{ at } r = R_0, \qquad (17)$$

where $N$ is the total oxygen transfer rate and $A$ is the surface area of the inner cylinder. Solving the axisymmetric diffusion equation in the space between the cylinders, subject to the boundary conditions Eqs (16) and (17), assuming axial diffusion to be negligible, gives the concentration $c_{\text{inner}}$ at the surface of the inner cylinder to be

$$c_{\text{inner}} = c_{\text{mat}} - \frac{R_0 N}{AD} \ln\left(1 + \frac{d}{R_0}\right) = c_{\text{mat}}(1 - N/N_{\text{max}}), \qquad (18)$$

where $N_{\text{max}}$ is the theoretical maximum for the oxygen transfer rate given by Eq (14). To increase the accuracy of the regression equation, the numerical values of $N_{\text{max}}$ given in Table 3 are used here. Now considering oxygen transfer into the capillary, taking the oxygen concentration at the capillary surface to be given by $c_{\text{inner}}$ from Eq (18), the oxygen transfer rate $N_{\text{flow}}$ from Eq (15) is multiplied by $c_{\text{inner}}/c_{\text{mat}}$ due to the linearity of the advection-diffusion equation. Using this correction with Eq (15) and rearranging gives a regression equation for the oxygen transfer rate $N$ accounting for the surrounding villous volume,

$$N \approx \frac{N_{\text{flow}}}{1 + N_{\text{flow}}/N_{\text{max}}}, \qquad (19)$$

where $N_{\text{flow}}$ is given by Eq (15) and $N_{\text{max}}$ is the upper bound on the oxygen transfer rate, calculated numerically. It is simple to adapt the regression Eq (19) to plot $N$ versus another quantity such as the flow rate $Q$ or the effective Péclet number $Pe_{\text{eff}}$.

The accuracy of the scalings Eqs (12) and (13) and the regression Eq (19) are tested across a range of Péclet numbers in Fig 3. For smaller values of $\Delta P$, where the scaling Eq (13) holds, the data of Fig 2A for each capillary collapse close to a single straight line when $NR$ is plotted against $\Delta P$ on log-log axes, as predicted (Fig 3A). For larger values of $\Delta P$, when $N/(L^2/R)^{1/3}$ is plotted against $\Delta P$ on log-log axes, the data collapse onto straight lines of approximate gradient 1/3, confirming that the scaling Eq (12) holds for the values of $\Delta P$ considered here. The regression curves given by Eq (19) qualitatively capture the numerical results across a range of values of $\Delta P$ (Fig 3C).

The numerical results obtained when the nonlinear oxygen–hemoglobin dissociation law is taken into account via replacing Eq (6) with Eq (4) are given in Fig 3D. The close agreement between the results in the nonlinear case and the linearized case can be seen by comparing Fig 3C and 3D. Also plotted in Fig 3D is the average enhancement to advection in the nonlinear

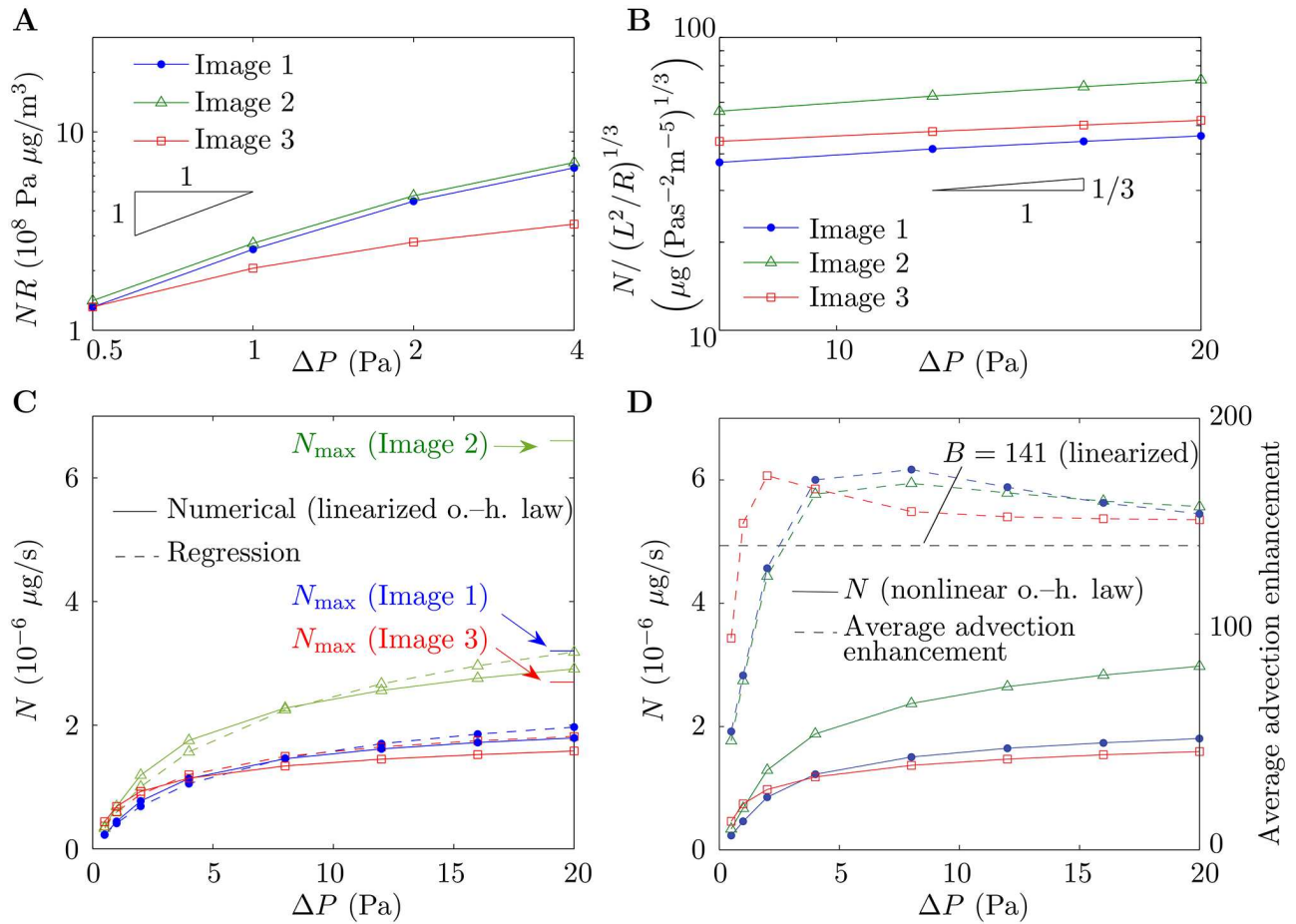

**Fig 3. Scaling and regression.** Throughout the figure, results for Image 1, Image 2 and Image 3 are denoted by blue dots, green triangles and red squares, respectively. (A,B) Scaled oxygen transfer rate (A) $NR$ and (B) $N/(L^2/R)^{1/3}$ versus pressure drop $\Delta P$, on log-log axes. Solid lines denote numerical results and black triangles denote the gradients predicted by (A) scaling Eq (13) and (B) scaling Eq (12). (C) Oxygen transfer rate $N$ versus $\Delta P$ with the linearized oxygen–hemoglobin dissociation law. Solid lines denote numerical results and dashed lines denote results generated by the regression Eq (19). The upper bounds $N_{max}$ on $N$ are indicated, which have been calculated by solving the diffusion equation in the surrounding villous volume with the capillary assumed to be deoxygenated. (D) Oxygen transfer rate $N$ with the nonlinear oxygen–hemoglobin dissociation law (solid lines) and average advection enhancement $1 + (c_{max}k_{hn}/\rho_{bl})S'(P_{O_2})$ in each capillary (dashed lines) versus $\Delta P$. Also shown on the figure is the oxygen advection enhancement parameter $B$ (black dashed line), derived by linearizing the oxygen–hemoglobin dissociation law.

doi:10.1371/journal.pone.0165369.g003

case, which depends on the pressure difference $\Delta P$, compared to the linearized value $B = 141$. The figure shows that the linearization overestimates the average advection enhancement at smaller values of $\Delta P$ and slightly underestimates the advection enhancement at larger values of $\Delta P$.

### Predictions of the idealized axisymmetric model and optimization of capillary dilation

There is a clear dilated bend in one of the branches of the capillary in Image 1 (Fig 1C). The size of the dilation can be seen in a plot of the radius of the capillary along its centreline, before it branches (Fig 4A). In this section an idealized model in an axisymmetric geometry (Figs 1E and 4D) is used to provide a possible physical explanation for this structure and to quantify its

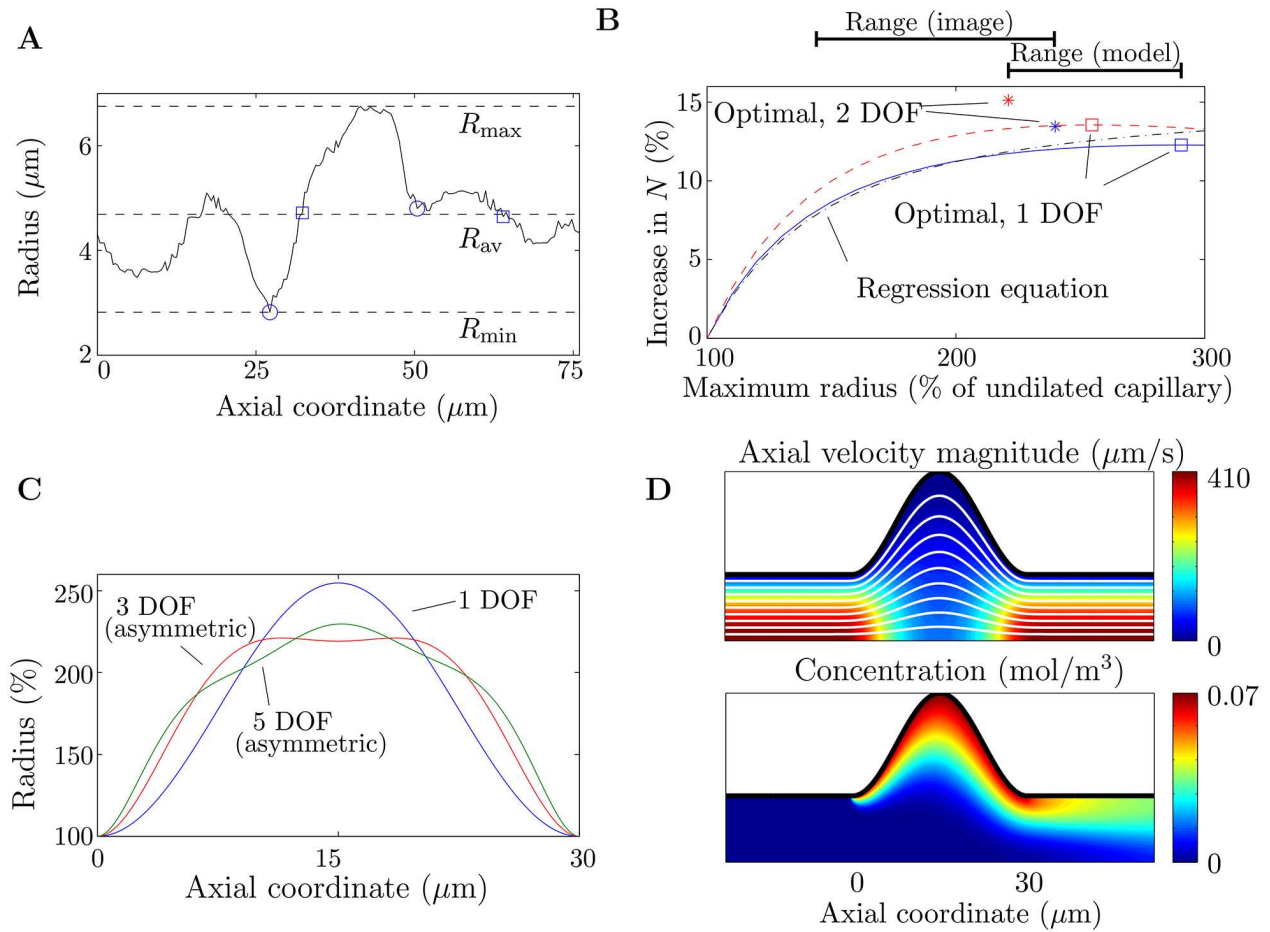

**Fig 4. Results from the idealized axisymmetric model of a fetal capillary dilation.** (A) Radius of the capillary (of length $L = 76$ μm) in Image 1 before it branches, showing the dilation and the different ways used to define its length $\lambda$. Local minima are denoted with circles (this is subjective due to noise in the data). Points where the radius of the dilation equals the average radius are denoted with squares. The dashed lines show the minimum radius $R_{\min} = 2.8$ μm, the average radius $R_{av} = 4.7$ μm and the maximum radius $R_{\max} = 6.8$ μm. (B) Increase in oxygen transfer rate for a capillary dilation compared to a straight capillary. Included in the figure are results for shapes with 1 degree of freedom (DOF), with the undilated radius being $R_{\min}$ ($\lambda = 23$ μm, $Pe_{\mathrm{eff}} = 285$; blue solid line) and $R_{av}$ ($\lambda = 31$ μm, $Pe_{\mathrm{eff}} = 475$; red dashed line). Indicated on the figure are the maximum radius and the increase in the oxygen transfer rate $N$ for the optimal shapes in each case with 1 DOF (squares) and 2 DOF (stars). The black dash-dotted line shows the predictions of the regression Eq (15), for a capillary dilation with 1 DOF, with the undilated radius defined as $R_{av}$. (C) Optimal shapes for successively increasing DOF. Where the shapes are asymmetric about the axial midpoint of the dilation, this is labeled on the figure. The shapes correspond to the case where the undilated radius of the capillary is defined as $R_{av}$, and are plotted as a percentage of the undilated radius. (D) Top: Axial velocity magnitude, with streamlines in white. Bottom: Concentration. Results are shown for the optimal dilation with 1 DOF, in the case where the undilated radius is taken to be $R_{av}$. It has been assumed that the centreline flow velocity in a straight capillary is equal to 300 μm/s.

doi:10.1371/journal.pone.0165369.g004

effect on oxygen transfer in the capillary. From a family of capillary dilations, we seek the shape that optimizes oxygen transfer over the dilated portion of the capillary.

For a capillary shape parameterized with one degree of freedom (DOF), an optimal maximum radius of the dilation is found, leading to the largest amount of oxygen entering the capillary (Fig 4B and 4C). Allowing an extra DOF in the shape of the dilation, while still imposing symmetry about the tube axis and the mid-plane of the dilation, gives a slightly smaller maximum radius of the optimal shape and an enhanced oxygen transfer rate (Fig 4B and 4C). Allowing further DOF in the shape of the dilation, including asymmetric shapes, does not reduce the maximum radius of the dilation (Fig 4C). The optimal shape increases the total oxygen transfer rate $N$ into the capillary by up to 15% above the transfer in a straight capillary (Fig 4B). The model predicts an optimal maximum radius of the dilation within approximately 33% of the radius of the dilation measured from Image 1 (Fig 4B). The regression Eq (15), with $\Delta P$ fixed, $L$ taken to be the length of the dilated section of the capillary, and the calculated values of the vascular resistance $R$ inserted, predicts the enhancement to $N$ in a dilated capillary with reasonable accuracy, although it does not capture the presence of an optimal maximum radius (Fig 4B). Due to the linearity of the governing equations with respect to concentration in the axisymmetric model, a change in the oxygen concentration difference between maternal and fetal blood does not affect the model predictions. The model predictions are not affected significantly by accounting for additional oxygen transfer through the villous volume upstream and downstream of the localized capillary dilation (see S3 Appendix).

The optimal maximum radius of the localized dilation arises due to two competing effects. When the size of the dilation increases, the overall resistance of the capillary is reduced, leading to an increased flow rate (pressure drop being fixed) and therefore an increase in oxygen transfer; the increased surface area also enhances transfer. However, by conservation of mass, the local flow speed reduces where the capillary radius increases, leading to a decrease in oxygen transfer (Fig 4D). If the maximum radius of the dilation is too large, the local slowdown in the flow speed overwhelms the increased flow rate and oxygen transfer decreases overall.

## Discussion

To capture the inherent stochastic variability in placental structure will require sampling of numerous villous branches to obtain distributions of placental geometries. This study demonstrates how scaling arguments, 3D simulations and idealized modeling can be combined to encapsulate key measures of flow resistance and nutrient transfer. The relevant flow characteristics that have been computed are the vascular resistance $R$ and the vascular resistance per unit villous volume $R_{vil}$ (Table 3). Using $R$, each capillary can be coupled to a model of the feto-placental vasculature to capture the global flow characteristics through the network. Similarly, $R_{vil}$ can be used to calculate total resistances through sections of a fetal villous tree of known volume.

We have applied scaling arguments to predict (at least for the three examples considered here) which capillaries are expected to receive the most oxygen from the maternal side across a range of Péclet numbers, using only the values for total capillary length and vascular resistance. The scaling laws, along with an upper bound on the oxygen transfer rate (which can be calculated by solving for diffusion in the villous volume with the capillary assumed to be deoxygenated), capture the relationship between $N$ and the pressure drop $\Delta P$ across parameter space where the effective Péclet number is large and linearized oxygen saturation kinetics can be assumed. These results are summarized in Table 1. The regression Eq (19) has been suggested as a way to calculate $N$ efficiently and has been shown to capture the numerical results qualitatively. The regression equation will be useful for estimating oxygen transfer in the terminal villi in future models of the entire placenta, where it will not be feasible to solve for $N$ numerically in every feto-placental capillary. Accurate values for the fetal oxygen transfer rate can be used to improve studies assuming the fetal villi to be oxygen sinks, such as [18, 19, 21, 23].

The identified scalings, given by Eqs (12) and (13), can also help quantify the effects of statistical variability and experimental uncertainty on blood flow and oxygen transfer. For example, tissue shrinkage or expansion can occur in villous branches for several reasons: first, due to a change in the difference between maternal and fetal blood pressure [49], for example as a result of a pathology such as pre-eclampsia; second, due to active constriction of venules or arterioles upstream and downstream of fetal capillaries, respectively [50]; finally, due to shrinkage during fixation [51]. The effect of tissue shrinkage or expansion on the data in Fig 2A has been determined by examining the dependence of the scalings Eqs (12) and (13) on the geometrical parameters. The numerical value of $N$ from Fig 2A can be multiplied by a factor depending on the magnitude of the shrinkage or expansion factor to recover the oxygen transfer rate; a similar technique has been used to correct structural quantities in stereological studies [51]. The sensitivity of the model to changes in the parameters can also be tested using Eqs (12) and (13), with the Formula (11) for the vascular resistance. Aside from the effects of changing the geometry discussed above, the scaling Eq (12) suggests that at lower values of $\Delta P$, the results are only affected by changes to $c_{mat}$, $B$ or $\mu$. At higher values of $\Delta P$, the results are expected to depend on all parameters, with the results being most sensitive to changes in the diffusion coefficient $D$.

The predictions of the model are sensitive to changes in specific aspects of placental structure, as a result of pathologies or differences between individuals. An idealized axisymmetric model has been used to investigate the effect of localized dilations in the feto-placental vasculature on blood flow and oxygen transfer. In the model, a localized dilation of optimal shape was found to increase oxygen transfer by up to 15%. Dilated capillaries are thought to occupy more than 35% of the total villous volume at term [5], and could therefore provide a significant enhancement to fetal oxygen uptake. The model supports the hypothesis that localized dilations develop towards term to enhance oxygen transfer in the placenta without wholesale placental growth or remodeling —in the second half of gestation, placental growth increases at a slower rate than fetal growth [1], while the mean trophoblast thickness decreases until term [52]. There may be some local remodeling of the capillary wall during the formation of the localized dilations, but we do not investigate the mechanism for their formation here. The enhancement to oxygen transfer could explain the increased dilation of fetal capillaries in pregnancies at high altitude [6, 8]; at 4300 m, maternal arterial partial pressure of oxygen is reduced by almost 50% compared to sea level [53]. A possibility is that healthy placentas at sea level are suboptimal, allowing expansion of dilations in hypoxic conditions. Increased vascularization of terminal villi has also been observed in placentas at high altitude [54], which could lead to an increase in the total number of capillary dilations. The results of the model could also provide a reason for the hypoxia-related outcomes of delayed villous maturation [14]. The lack of vasculo–scyncytial membranes observed in cases of delayed villous maturation could represent a lack of localized fetal dilations and therefore an inability of the fetus to extract an appropriate amount of oxygen from the maternal blood.

As well as characterizing the sensitivity of the model to changes in the parameters, the geometry and placental structure, it is important to understand the limitations of the modeling itself when interpreting the results. In both the 3D and axisymmetric models, simplifications have been made to the equations governing blood flow and oxygen transfer. The equations have a well-known asymptotic solution in certain limits of the effective Péclet number, allowing the scalings used in the paper to be identified. However, it will be important in future studies to relax some of the assumptions that were made in employing them. In particular, the non-Newtonian effects of red blood cells on blood flow in feto-placental capillaries should be investigated. Such effects are expected to be particularly important in small vessels and it will be vital to understand how the statistical distributions of quantities such as vascular resistance are affected by their inclusion in blood flow simulations. Non-Newtonian effects can often be incorporated into models using effective parameters or simplified equations (see [55], for example), but understanding whether such methods can be used in feto-placental capillaries of varying radius will necessitate a dedicated investigation involving simulations with discrete red blood cells.

Errors have also been introduced in converting the 3D geometries to finite-element meshes, via the neglect of the glycocalyx and by the smoothing operations performed prior to conversion, which will lead to a loss of detail in the local topography. A limitation of the current study is that the 3D images were obtained from placentas that were not perfusion-fixed. The vascular resistances calculated in this study could therefore be overestimated. It is thought that the expansion of fetal capillaries with increasing arterial blood pressure could be especially prevalent in localized dilations [56], which could mean the radii of localized dilations are particularly underestimated in Fig 4A.

To summarize, in this paper we have investigated blood flow and oxygen transfer in feto-placental capillaries theoretically, using a combination of 3D simulations, physical scaling arguments and an idealized axisymmetric model. Simulations have been performed on real geometries by converting 3D images of villous and capillary surfaces to finite-element meshes. The relationship between the total oxygen transfer rate and the pressure drop through the capillary has been captured across a wide range of pressure drops by physical scaling laws and an upper bound on the oxygen transfer rate, which has been linked directly to the geometrical parameters. A regression equation has been introduced to estimate the total oxygen transfer using the vascular resistance. Scaling arguments have been shown to be an efficient tool for quantifying the effect of statistical variability and experimental uncertainty; in particular, the identified scalings have been used to estimate the effects of tissue shrinkage or expansion in villous branches on blood flow and oxygen transfer. An idealized model in an axisymmetric geometry has been used to quantify the effect of localized capillary dilations in the fetal vasculature on oxygen transfer. The model predictions support the hypothesis that localized dilations enhance oxygen transfer towards term without the need for extensive placental growth or remodeling. The combination of 3D simulations, scaling arguments and idealized modeling used in the paper provides a blueprint for quantifying variation in measures of flow resistance and oxygen transfer as further 3D images of feto-placental villous trees become available.

## Supporting Information

**S1 Images. Three-dimensional images.** A zipped folder containing the 3D surface meshes imported into *Comsol Multiphysics* for the simulations (.stl files, which can be opened in *Meshlab*) and the 3D tetrahedral meshes used for capillary skeletonization and calculation of the average villous distances (.bdf files, which can be opened in *Gmsh*).
(ZIP)

**S1 Appendix. Methods used for image analysis and three-dimensional simulations.** A description of the methods used in the image analysis and 3D simulations to increase the reproducibility of the results.
(PDF)

**S2 Appendix. Estimation of the oxygen advection enhancement parameter *B*.** The derivation of the parameter *B* quantifying the enhancement to oxygen advection due to its affinity for hemoglobin.
(PDF)

**S3 Appendix. Investigation of the effect of oxygen transfer upstream and downstream of a localized dilation.** Further numerical results generated using the idealized axisymmetric model, to quantify the effect of oxygen transfer through the surrounding villous volume upstream and downstream of the localized dilation.
(PDF)

## Acknowledgments

We acknowledge helpful discussions with Simon Cotter, Ian Crocker, Alexander Heazell and Edward Johnstone.

## Author Contributions

**Conceptualization:** PP PB OEJ.

**Data curation:** PP.

**Formal analysis:** PP ILC OEJ.

**Funding acquisition:** PP ILC OEJ.

**Investigation:** PP.

**Methodology:** PP ILC OEJ.

**Project administration:** PP OEJ.

**Resources:** JJ MJ LK.

**Software:** PP.

**Supervision:** PB OEJ.

**Validation:** PP.

**Visualization:** PP.

**Writing – original draft:** PP PB OEJ.

**Writing – review & editing:** PP PB JJ MJ LK ILC OEJ.